\documentclass[%
 reprint,
nobibnotes,
amsmath,amssymb,
aps,
prl,
]{revtex4-1}

\usepackage{graphicx}
\usepackage{xcolor}
\usepackage{braket}
\usepackage{siunitx}

\begin{document}

\title{On-Demand Correlated Errors in Superconducting Qubits from a Particle Accelerator}%

\author{Thomas McJunkin}\email{tom.mcjunkin@jhuapl.edu}
\author{A.W. Hunt}
\author{Yenuel Jones-Alberty}
\author{T.M. Haard}
\author{M.K. Spear}
\author{James Shackford}
\author{Tom Gilliss}
\author{Mayra Amezcua}
\author{C.A. Watson}
\author{T.M. Sweeney}
\author{J.A. Hoffmann}
\author{Kevin Schultz}

\affiliation{Johns Hopkins Applied Physics Laboratory, Laurel, MD 20723, USA}

\date{\today}

\begin{abstract}
Ionizing radiation is a known source of correlated errors in superconducting quantum processors, inhibiting the functionality of quantum error correction surface codes. High-energy photons and charged particles deposit pair-breaking energy into these systems leading to excess quasiparticles near Josephson junctions that increase qubit decoherence. Previous investigations of this problem have relied on ambient, stochastic sources of ionizing radiation or alternative methods of quasiparticle generation. Here, we present a facility that couples an electron linear accelerator (linac) to a dilution refrigerator to study ionizing radiation in quantum systems. A single linac electron closely mimics the energy deposition characteristics of a typical cosmic-ray muon, and we demonstrate the facility's usefulness with a multi-qubit superconducting transmon chip. Characteristic radiation-induced relaxation errors are quickly and easily collected with the speed and timing information of the linac. Additionally, we present qubit excitation and detuning errors that can be difficult to detect without the on-demand source of ionizing radiation. These error signatures are shown to be dependent on the junction placement and surrounding superconducting gaps.
\end{abstract}

\maketitle

\textit{Introduction\textemdash}Superconducting circuits containing Josephson junctions are a leading qubit platform~\cite{blaisCircuitQuantumElectrodynamics2021}, with below-threshold quantum error correction demonstrated in multiple systems~\cite{acharyaQuantumErrorCorrection2024, sivakRealTimeQuantumErrorCorrection2023}. 
Among the myriad technological hurdles on the path to a useful quantum computer, superconducting qubits are particularly vulnerable to decoherence caused by ionizing radiation~\cite{martinisSavingSuperconductingQuantum2021}. 
High-energy charged particles (e.g. muons, electrons) transit materials in tracks of ionization and subsequent electron-hole recombination that release bursts of ballistic phonons into the substrate (typically Si), some of which enter the superconducting layer where they break Cooper pairs into quasiparticles~(QPs)~\cite{linehanEstimatingEnergyThreshold2025}. 
There, the excess QPs can reach the Josephson junction of the qubit and tunnel across, resulting in qubit decoherence~\cite{lenanderMeasurementEnergyDecay2011, catelaniRelaxationFrequencyShifts2011, catelaniDecoherenceSuperconductingQubits2012}. 
With the chip-scale spread of phonons, a single ionizing particle can substantially increase decoherence across many qubits for a computationally long time.
Though shielding and material purification can mitigate most sources of ionizing radiation, atmospheric muons are the primary exception due to their unusually high penetration strength~\cite{loerAbatementIonizingRadiation2024, cardaniReducingImpactRadioactivity2021}. 
In the past half-decade, high-energy particle impacts have been identified as a leading cause of correlated qubit errors threatening prolonged, reliable operation of quantum error correction schemes~\cite{vepsalainenImpactIonizingRadiation2020, wilenCorrelatedChargeNoise2021, cardaniReducingImpactRadioactivity2021, mcewenResolvingCatastrophicError2022, thorbeckTwoLevelSystemDynamicsSuperconducting2023,
larsonQuasiparticlePoisoningSuperconducting2025, kurilovich2025correlatederrorburstsgapengineered}. 

Studies of particle impacts on superconducting qubits have often relied on ambient, stochastic sources of ionizing radiation. This limits the timing resolution of the impact, and results in costly data collection times for sufficient event statistics. Some recent studies have utilized coincident detection schemes to improve the timing resolution~\cite{harringtonSynchronousDetectionCosmic2024, liDirectEvidenceCosmicrayinduced2024a}, but this approach is only useful for certain particle types and does not solve the event-frequency concern. Others have turned to alternative QP generation techniques such as optical illumination or junction biasing to circumvent the issues of ambient radiation, but the exact QP dynamics of these approaches may not faithfully recreate the effects of an ionizing particle~\cite{iaiaPhononDownconversionSuppress2022, yeltonModelingPhononmediatedQuasiparticle2024, mcewenResistingHighEnergyImpact2024, benevidesQuasiparticleDynamicsSuperconducting2024}. 

Here, we present a new experimental facility for studying the effects of high-energy radiation on cryogenic quantum systems: CLIQUE (Controlled Linac Irradiation of Quantum Experiments)\footnote{CLIQUE.Facility@jhuapl.edu}. The pulsed beam from an electron linear accelerator (linac) is directed at a multi-qubit transmon chip, depositing energy from a single 18.5~\si{\mega \electronvolt} electron with \textless$10$~\si{\micro \second} precision. The result is an on-demand and heralded source of high-energy radiation that closely mimics the energy deposition of a single cosmic-ray muon. We present details of the facility and its operation, as well as comparisons between the linac electrons and muons. We show how individual collision events can be easily identified and averaged for overall system and qubit-specific behavior. We further highlight the value of this on-demand scheme by presenting more subtle qubit excitation errors as well as detuning and dephasing errors. We note that junction details and junction orientation in relation to the capacitor island in these grounded transmons greatly affect the various error manifestations. We conclude with a discussion of future research directions of the CLIQUE facility. 

\begin{figure*}[t]
\centering
\includegraphics[width = 0.98\textwidth]{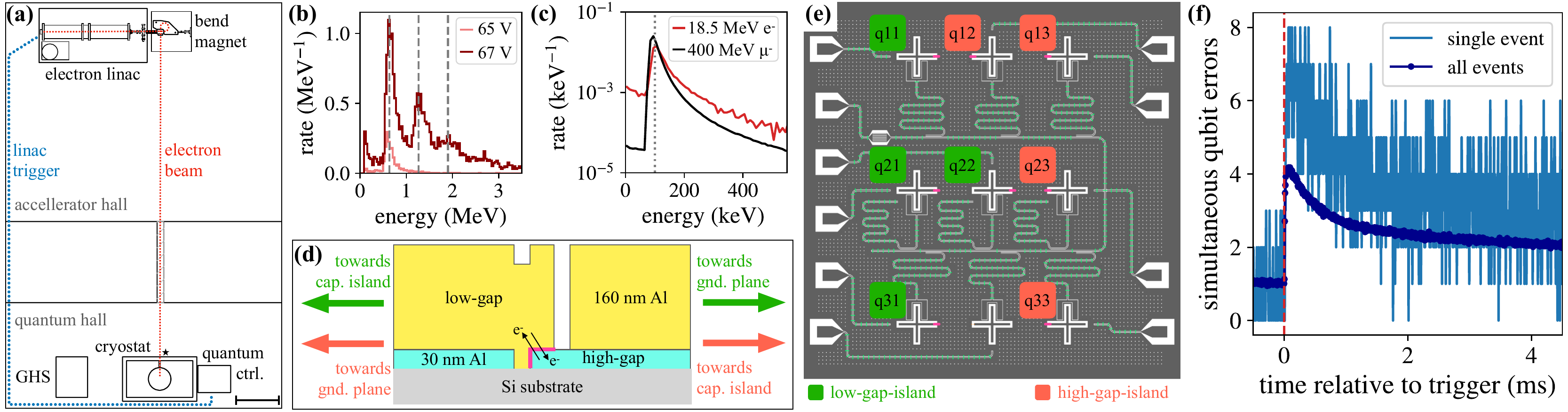}
\caption{
\textbf{Experimental Design.} (a)~Schematic of the CLIQUE facility. The electron beam (red dashed line) exits the linac, redirects through the bending magnet, and travels into the adjoining room where it strikes the dilution refrigerator. The linac trigger (blue dotted line) is supplied to both the linac and the quantum control hardware for event tagging. A scintillation detector (star) is positioned just off-axis from the incident beam to continuously monitor the electron fluence. The scale bar (lower right) is 1~\si{\meter}. (b)~Scintillator response to electron fluence tuning. The probability distribution of energy deposition in the scintillator is collected for two linac gun grid voltage pulses. Vertical gray dashed lines mark the expected energy deposition for 1, 2, and 3 electrons. (c)~MCNP simulation comparison of the energy deposition distribution for linac electrons and typical cosmic-ray muons in a 350~\si{\micro \meter} silicon chip. The most-probable energy deposition, 100~\si{\kilo \electronvolt}, is marked with a vertical gray dashed line. 
(d)~Josephson junction design, indicating the superconducting gaps of the layers and their orientations relative to the capacitor island or ground plane for the two layouts (green vs. red labeled qubits). The junction interface is shown in pink, where QPs (unpaired electrons) can tunnel across and cause qubit errors. (e)~Design schematic of the qubit chip, with labeled qubits and junction locations highlighted in pink. (f)~Relaxation detection measurement. All qubits are prepared in the $\ket{1}$ state with a 1~\si{\micro \second} detection delay before measurement. The lighter blue curve is a single correlated error event coincident with the linac trigger and the darker blue points are the average of all 831 detected linac events across $\sim$7.4~\si{\minute} of data.
}
\label{fig:intro}
\end{figure*}

\textit{Quantum Radiation Facility\textemdash}Figure~\ref{fig:intro}(a) presents a schematic of the CLIQUE facility. The linac and dilution refrigerator are housed in neighboring rooms separated by 1.8~\si{\meter} of reinforced concrete. The electrons are sourced in a thermionic gun pulsed at 10~\si{\hertz} for 4~\si{\micro \second}, accelerated by 2856~\si{\mega \hertz} RF power through a re-purposed Varian Clinac-2500 resonant cavity structure, and exit the linac with 18.5~\si{\mega \electronvolt} of kinetic energy. The electron beam is redirected through a 270\si{\degree} achromatic bending magnet, then exits to air and diverges as it travels 4.5~\si{\meter} across the linac room and through a 14.6~\si{\centi \meter} diameter hole in the concrete wall. Once in the quantum hall, the electrons travel an additional 1.3~\si{\meter} where they impact the outer shield of the dilution refrigerator. All shield layers have 3~\si{\centi \meter} optical access windows positioned at the sample space below the mixing chamber plate. The windows, titanium on the vacuum shield and aluminum on all other shields, are 25~\si{\micro \meter} thick to minimize energy loss and secondary generation of the incident electrons. The device packaging is additionally thinned in the beam path in front of the qubit chip to 800~\si{\micro \meter} of aluminum.

The electron fluence is continuously monitored by a scintillation detector positioned a few cm off axis from the windows. Figure~\ref{fig:intro}(b) shows a test of the linac electron fluence tunability. The deposited energy in the detector shows peaks corresponding to integer electron counts, tunable with the linac gun grid voltage. For this work, the fluence was tuned to $\sim$0.2~electrons per pulse at the qubit chip to ensure no more than one electron for the vast majority of pulses ($\sim$98\si{\percent}).

Muons and other relativistic charged particles are often minimum-ionization particles~(MIPs), so-called because their energy losses in a given material are similar and lie within the broad basin of minimum ionization loss described by the Bethe-Bloch equation~\cite{PhysRevD.110.030001}. Electrons with kinetic energy in the range of 10-20~\si{\mega \electronvolt} passing through silicon also lose energy chiefly through ionizing collisions~\cite{mjberger2005}. Figure~\ref{fig:intro}(c) compares the energy-deposition probability for 18.5~\si{\mega \electronvolt} electrons and 400~\si{\mega \electronvolt} muons passing through 350~\si{\micro \meter} of silicon, showing remarkable similarity~\cite{rastin1984}. Thus, linac electrons in the CLIQUE facility serve well as surrogates for the energy-deposition characteristics of MIPs generally and of muons specifically. Further details about the facility are provided in the supplemental materials.

\textit{Correlated Error Experiment\textemdash}The quantum system in this study is a superconducting qubit chip containing nine aluminum fixed-frequency transmon qubits.
The grounded transmon design and Dolan bridge construction of the Josephson junctions lead to a difference in qubit design dependent on the location of the junction~\cite{dolanOffsetMasksLiftOff1977}. Figure~\ref{fig:intro}(d) presents a schematic of the junction, highlighting the differences in layer thickness which results in a superconducting gap difference~($\sim$2.8~\si{\giga \hertz}) across the junction.
Eight qubits, labeled as ``q\#\#" in Fig.~\ref{fig:intro}(e), are controlled with independent voltage lines and simultaneously measured through a shared transmission line. Qubits labeled green have the lower gap side of the junction connected to the capacitor island (``low-gap-island"), while red-labeled qubits have the higher gap side connected to the capacitor island (``high-gap-island"). As previously reported, this incidental gap engineering can lead to different responses to QP poisoning events~\cite{harringtonSynchronousDetectionCosmic2024}.

The event detection scheme is comprised of four components: preparation, detection delay, measurement, and reset. A qubit is determined to have errored if not measured to be in the expected prepared state. The reset step conditionally applies a $\pi$ rotation if the measurement returns the excited state. The trigger pulse sent to the linac is additionally routed to the quantum control system [see Fig.~\ref{fig:intro}(a)], providing a time stamp of electron incidence, to within the 4~\si{\micro \second} pulse width of the linac gun. Practically, the trigger is assigned to a single quantum shot, which varied from $\sim$5 to 15~\si{\micro \second} depending on detection delay. Figure~\ref{fig:intro}(f) shows a typical system response to a linac pulse when all qubits are prepared in the $\ket{1}$ state, which is sensitive to relaxation errors. The baseline of $\sim$1 qubit error sharply rises to full system error immediately following the linac trigger, then slowly returns to the baseline. Because the electron fluence is much less than one per pulse, a simple event classification scheme is used to identify collision events correlated with the linac trigger, see supplemental materials for details. With trigger timing information, all events can easily be combined to reconstruct an averaged system response: the dark blue curve in Fig.~\ref{fig:intro}(f). The short- and long-timescale exponential decays are a result of the two junction orientations present on the device. 

\begin{figure}
\centering
\includegraphics[width = 0.49\textwidth]{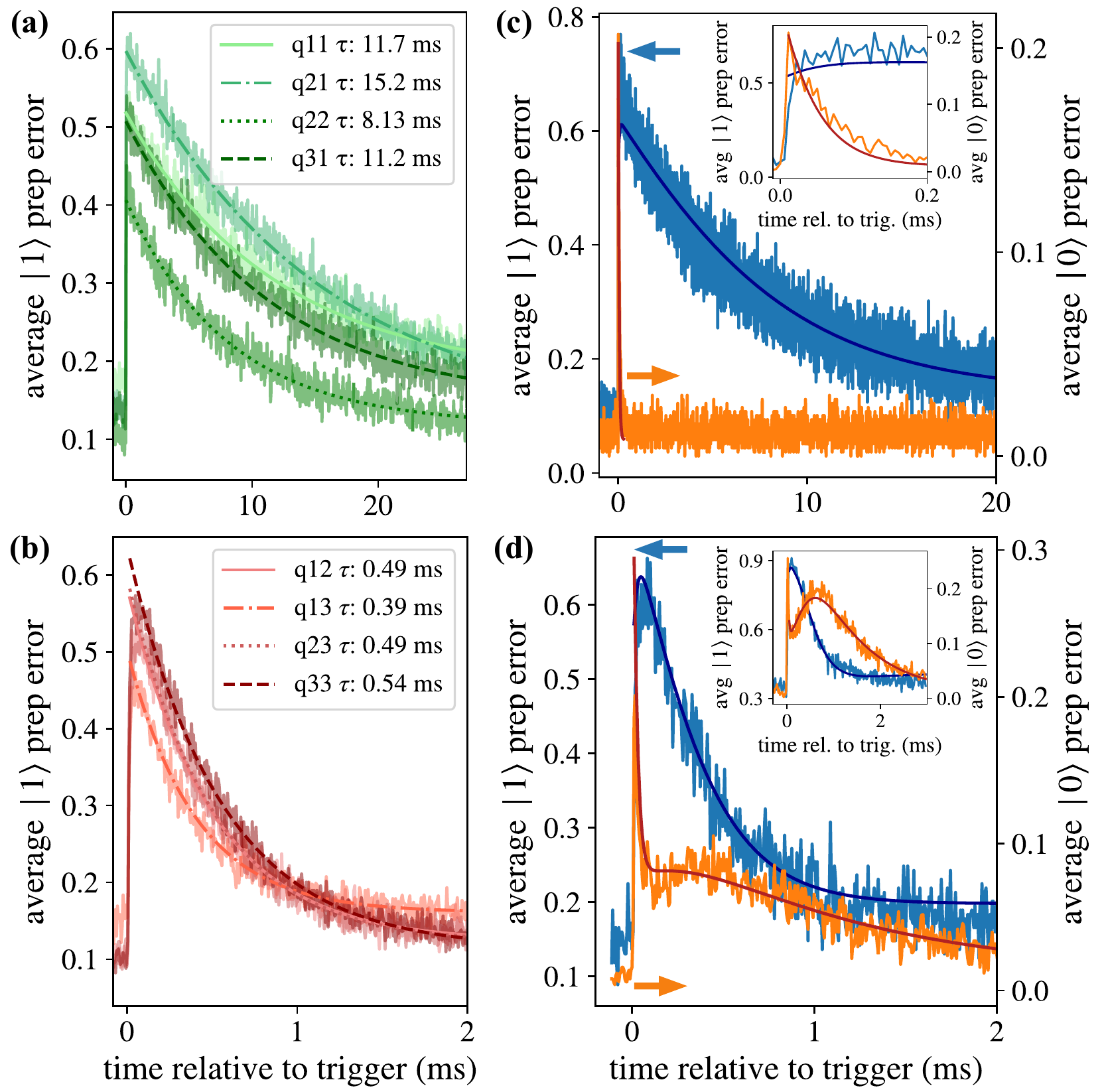}
\caption{
\textbf{Design-dependent relaxation and excitation errors.} (a)~Error signature for $\ket{1}$ state preparation with 1~\si{\micro \second} detection delay averaged across all detected linac events for low-gap-island qubits [green in Fig.~\ref{fig:intro}(e)]. Narrow lines are exponential fits ($A e^{-t/\tau} + c$) for each qubit. (b)~Data collected coincidentally with that of (a) and analyzed similarly, for high-gap-island qubits [red in Fig.~\ref{fig:intro}(e)]. (c)~Error signature for $\ket{1}$ state prep (blue) and $\ket{0}$ state prep (orange) with 1~\si{\micro \second} detection delay across the largest 20\si{\percent} of detected linac events for q22 (low-gap-island). The darker narrow curves are fits using the rate equation dynamics described in the main text. The inset shows the same data and fits in a shorter time window following the linac trigger. (d)~Data collected coincidentally with that of (c) and analyzed similarly for q23 (high-gap-island). The inset shows the q23 results for a 12~\si{\micro \second} detection delay using the same fit parameters. 
}
\label{fig:prep10}
\end{figure}

\textit{Detailed Error Behavior\textemdash}While whole-system response is used for event discrimination, the error responses of all qubits are saved separately for individual analysis. Figures~\ref{fig:prep10}(a-b) show the averaged error response of all eight qubits for $\ket{1}$ state preparation and 1~\si{\micro \second} detection delay. Low-gap-island qubits [Fig.~\ref{fig:prep10}(a)] show at least an order of magnitude longer recovery time to baseline error than the oppositely oriented qubits [Fig.~\ref{fig:prep10}(b)]. This can be explained by the following three assumptions. First, QP density in a higher-gapped region should be exponentially suppressed compared to a low-gap region~\cite{marchegianiQuasiparticlesSuperconductingQubits2022}. Second, QP flow in the low-to-high gap direction is expected to be the primary route for relaxation-induced QP tunneling~\cite{connollyCoexistenceNonequilibriumDensity2024a}. Lastly, due to the larger area, QP trapping features, and wire-bond escape routes, QP densities and energies should be lower in the ground plane than on the capacitor island~\cite{harringtonSynchronousDetectionCosmic2024}. In combination, these factors led to the longer-lived relaxation errors in low-gap-island qubits than high-gap-island qubits.

QP tunneling can also excite qubits and preparing the $\ket{0}$ state exposes these excitation errors, which are similarly influenced by the junction orientation and gap differences. Figures~\ref{fig:prep10}(c-d) show the error response for $\ket{1}$ state preparation (blue) and $\ket{0}$ state preparation (orange) with 1~\si{\micro \second} detection delay for example low-gap-island and high-gap-island qubits, q22 and q23, respectively. Only collision events in the top 20\si{\percent} as measured by the event detection scheme are shown due to the significantly smaller error signatures for $\ket{0}$ state preparation. For q22, the excitation errors are only briefly visible in the first $\sim$100~\si{\micro \second} [see Fig.~\ref{fig:prep10}(c) inset], whereas q23 shows additional longer-lived excitation errors. As excitation errors are expected to be primarily driven by QP tunneling in the high-to-low gap direction~\cite{connollyCoexistenceNonequilibriumDensity2024a}, the observed behavior in q23 can be explained by prolonged QP density on the capacitor island. The brief spike in excitation error shown in both junction orientations is likely due to out-of-equilibrium QPs not yet cooled to the superconducting gap leading to enhanced tunneling and energy exchange rates~\cite{marchegianiNonequilibriumregimes2025}.

A phenomenological two-level rate model was used to better understand the time dynamics of the error signatures in Fig.s~\ref{fig:prep10}(c-d). The qubit state dynamics are represented by the coupled two-state linear differential equation
\begin{equation}
\frac{dP(t)}{dt} = 
\begin{pmatrix}
-\Gamma_{exc} && \Gamma_{rel} + \Gamma_{T_1} \\
\Gamma_{exc} && -\Gamma_{rel} - \Gamma_{T_1}
\end{pmatrix} P(t).
\label{eq:Pt}
\end{equation}
Here, $t$ is time, $\Gamma_{T_1}$ is the static $T_1$ relaxation rate, and $P = \begin{pmatrix} P_{\ket0} \\ P_{\ket1}  \end{pmatrix}$. The time-dependent relaxation and excitation rates $\Gamma_{rel} = A_{rel}e^{-t/\tau_{rel}}$ and $\Gamma_{exc} = A_{exc_1}e^{-t/\tau_{exc_1}} + A_{exc_2}e^{-t/\tau_{exc_2}}$ are assumed to exponentially decay from the linac trigger time $(t = 0)$ as the QPs diffuse. For high-gap-island qubits [Fig.~\ref{fig:prep10}(d)], the excitation rate $\Gamma_{exc}$ contains two time-dependent components to capture the short- and long-time signatures in the data. The second term is excluded for low-gap-island qubits. 

The differential equation in Eq.~\ref{eq:Pt} was solved over the time of a single shot with an ordinary differential equation solver to obtain a final state after time-evolution under these rates. This was repeated as the initial time is stepped from the linac trigger and fit across all time steps to extract the amplitudes, $A$, and decay constants, $\tau$. Independent measurements of $T_1$ interleaved between measurement runs were used to estimate $\Gamma_{T_1}$ and used to create an appropriate preparation/measurement correction matrix applied at the beginning and end of integration, see supplemental materials for details. The fit includes data from seven different detection delays, spanning 200~\si{\nano \second} to 12~\si{\micro \second}.

\begin{table}[h]
\caption{\label{tab:fitdata}\textbf{Fit parameters for Fig.~\ref{fig:prep10}(c-d)}. Amplitudes, $A$, are listed in ~\si{\kilo \hertz} and decay times, $\tau$, are listed in ~\si{\micro \second}. Reported uncertainties are standard deviations from the fit. Fit results for all eight measured qubits can be found in the supplemental materials.}
\begin{ruledtabular}
\begin{tabular}{ccccccc}
qubit & $A_{rel}$ & $\tau_{rel}$ & $A_{exc_1}$ & $\tau_{exc_1}$ & $A_{exc_2}$ & $\tau_{exc_2}$ \\
\hline
q22 & 272$\pm$1 & 6240$\pm$10 & 162$\pm$8 & 42$\pm$2 & N/A & N/A \\
q23 & 611$\pm$5 & 370$\pm$2 & 46$\pm$1 & 960$\pm$20 & 760$\pm$50 & 17$\pm$1 \\
\end{tabular}
\end{ruledtabular}
\end{table}

The fits for q22 and q23 are shown as darker blue and red lines in Fig.~\ref{fig:prep10}(c-d) and fit parameters are listed in Table~\ref{tab:fitdata}. The inset to Fig~\ref{fig:prep10}(d) shows the same fit parameters for q23 overlaid on the 12~\si{\micro \second} detection delay data, showing good agreement across detection delays. With this model, the initial rounding of the relaxation error can be explained by a brief but strong excitation error while the dip in excitation error for high-gap-island qubits around 200-300~\si{\micro \second} is due to the strong relaxation errors limiting the visibility of excitation errors. Similarly prolonged excitation errors may exist for low-gap-island qubits, but the extremely long-lived relaxation errors would strongly dampen their visibility. Of note, the longer excitation decay time for q23 is $\sim$3X longer than the relaxation decay time. Since QP-induced excitation and relaxation are primarily driven by tunneling in opposite directions, this result gives insight into the location dependence of QP density for this qubit. Modeling that considers the various junction layers (gaps, volumes, locations), QP dynamics (diffusion and recombination rates), and higher transmon energy levels may improve these fits but was not the focus of this work~\cite{pinckney_inprep}.

\begin{figure}[t]
\centering
\includegraphics[width = 0.49\textwidth]{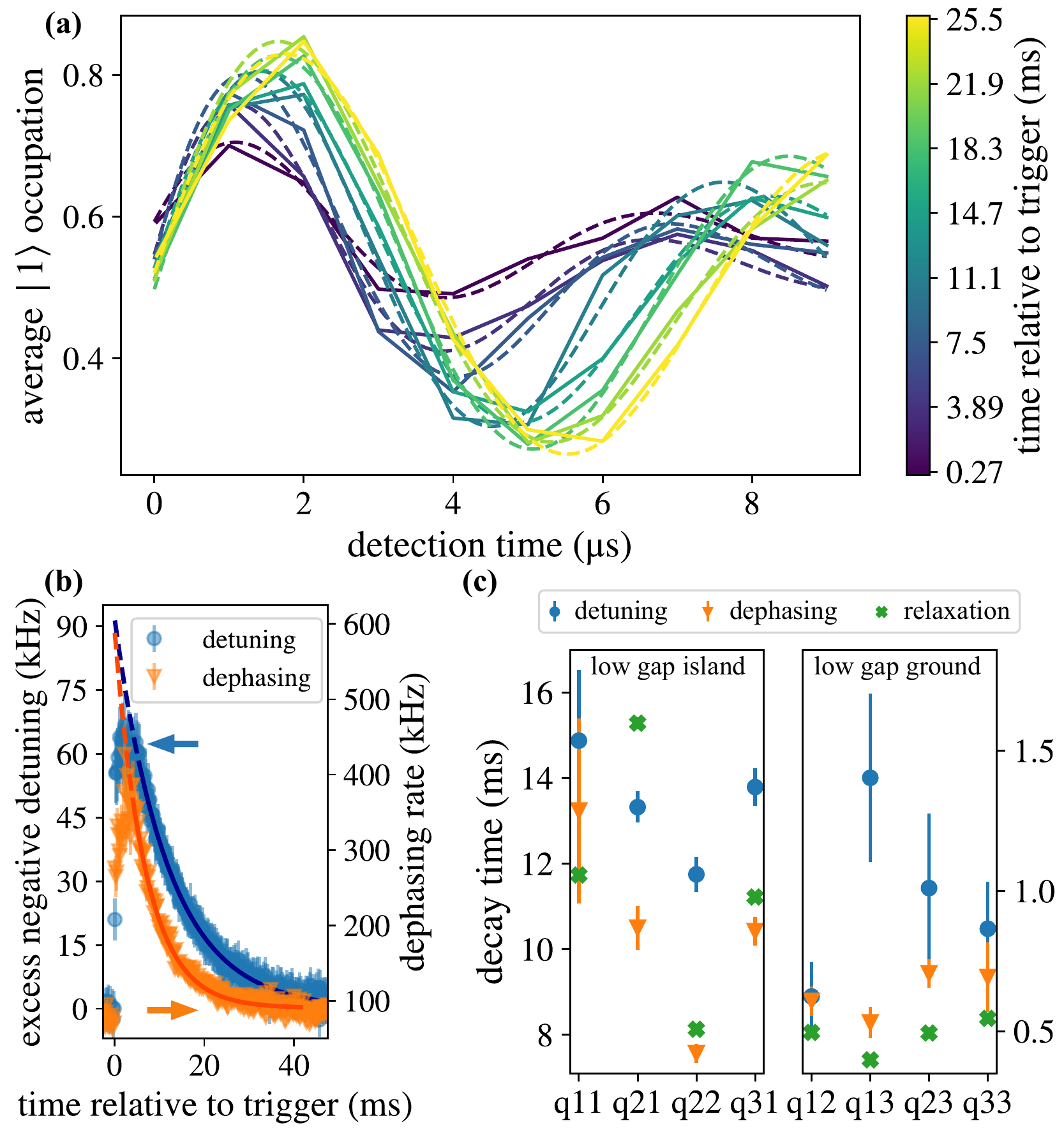}
\caption{
\textbf{Detuning and dephasing errors.} (a)~Time slices following the linac trigger of the Ramsey-like detection scheme for q22. Solid curves are the averaged data across 542 detected events and dashed lines are decaying sinusoidal fits ($A e^{-t/T_\phi} \sin{(\omega t_{detect} + \phi_0)} + c$). (b)~Extracted  detuning (blue circles) and dephasing rates (orange triangles) for q22. The plotted detuning excludes the intentional offset of the Ramsey experiment. Solid lines are exponential fits to the data (dashed to extend back to the trigger time). (c)~Comparison of detuning, dephasing, and relaxation error recovery times for all qubits. Error bars in (b)and (c) are standard deviations from the fits.
}
\label{fig:detuning}
\end{figure}

In addition to tunneling-dependent relaxation and excitation errors, high QP density near qubit junctions reduces their critical currents, negatively detuning the transmon frequency~\cite{catelaniRelaxationFrequencyShifts2011, wangMeasurementcontrolquasiparticle2014, kurilovich2025correlatederrorburstsgapengineered}. To measure detuning in response to linac electrons, the qubits were prepared along the Y-axis, allowed to precess for a varying amount of time during the detection delay, then projected to the X-axis for measurement. The qubit control pulses were intentionally detuned by +100~\si{\kilo \hertz} to effectively detune the qubit by -100~\si{\kilo \hertz}. This Ramsey-like sequence was repeated and aligned in post-processing to the linac trigger to create the averaged oscillating data shown in Fig.~\ref{fig:detuning}(a) for q22. Following the linac trigger, the oscillation decay time and frequency are reduced, then slowly recover. The dashed lines in Fig.~\ref{fig:detuning}(a) are decaying sinusoidal fits to the data. The resulting fit decay constant and frequency (removing the intentional offset) are shown in Fig.~\ref{fig:detuning}(b). The reduced detuning and dephasing immediately following the linac trigger are likely due to poor fits from strong relaxation effects and the limited number of detection delay points, so exponential fits to the data in Fig.~\ref{fig:detuning}(b) exclude this region. 

Figure~\ref{fig:detuning}(c) compares the decay times for these Ramsey measurements with the results from Fig.~\ref{fig:prep10}(a-b). For both junction orientations, the average detuning error persists for much longer than relaxation errors. As the detuning effect is not dependent on QP energy or their tunneling across the junction, these errors will persist as long as there is elevated QP density on either side of the junction. It is important to note, however, that the high-gap-island qubits show quicker recovery for all error manifestations. This highlights the importance of flushing QPs off the island and away from the junction for QP mitigation strategies~\cite{pinckney_inprep}.

\textit{Discussion\textemdash}There are several aspects of the CLIQUE facility that make it a powerful and enabling resource for studying ionizing radiation effects in quantum systems. Most importantly, the timing precision of the electron irradiation greatly simplifies collision detection and allows for averaging away other noise sources. Relaxation time scales are made apparent, and the more subtle impacts like qubit detuning are readily revealed, accurately tracked, and free of potential masking by shortened relaxation times. Secondly, the high rate of the linac enables gathering sufficient statistics on device behavior in drastically reduced time compared to ambient radiation sources (minutes versus hours, days, or weeks). Lastly, due to the similarity of energy deposition for a single linac electron and a cosmic-ray muon, the observed effects are an accurate reproduction of the response to critical ambient particle sources. 

Leveraging the advantages of the CLIQUE facility, we have shown that single electrons from the linac result in correlated relaxation, excitation, and detuning errors across several transmon qubits, highlighting the multiple pathways for ionizing radiation to cause qubit errors. Additionally, the differences observed in magnitudes and lifetimes for the error manifestations and junction orientations warrant further exploration of the rich QP dynamics present within a transmon qubit. Beyond the experiments presented here, there are many opportunities for future ionizing radiation studies with this facility such as two-level system dynamics, charge-parity measurements, effects from changes in magnetic flux, higher energy regimes through multi-electron depositions, and other quantum platforms. As a final note, we invite collaborations with interested researchers to utilize these resources and advance our shared understanding of ionizing radiation impacts in quantum systems. 

\hfill

\textit{Acknowledgments\textemdash}The qubit device used in this study was fabricated at
MIT Lincoln Laboratory under the program Superconducting Qubits at Lincoln Laboratory (SQUILL). The authors acknowledge JHU/APL for the infrastructure investments required to perform these experiments.

\bibliography{pew_paper_bib_v2}



\clearpage
\newpage
\maketitle
\onecolumngrid
\begin{center}

{ \Large\bf Supplemental Material}
\end{center}

\setcounter{equation}{0}
\setcounter{figure}{0}
\setcounter{section}{0}
\setcounter{table}{0}

\section{CLIQUE Facility}

\subsection{Electron Linear Accelerator}

The high-energy electrons were generated by a pulsed standing-wave radio frequency linear accelerator (linac). The accelerating structure was repurposed from a Varian Clinac-2500 radiation therapy system~\cite{tanabe81}. Figure~\ref{fig:supp_facility}(a) presents a picture of this linac. The electron beam starts at the thermionic triode electron gun with a cathode potential of -18~\si{\kilo \volt}. When the gun's grid potential is raised above the cathode potential, electrons are injected into the accelerating waveguide. The electromagnetic fields in the structure's resonant cavities are excited by a 10~\si{\micro \second} wide pulse of 2856~\si{\mega \hertz} radio frequency with a few megawatts of peak power. There are a total of 36 accelerating cavities, with each cavity imparting approximately 510~\si{\kilo \electronvolt} of kinetic energy for a total of 18.5~\si{\mega \electronvolt} at the end of the $\sim$1.9~\si{\meter} long waveguide. The electrons then drift $\sim$84~\si{\centi \meter} to a 270\si{\degree} achromatic bending magnet with a $\sim$15~\si{\centi \meter} effective bending radius and a set of copper slits at $\sim$90\si{\degree} to limit the energy spread to $\pm$1.5\si{\percent}, before directing the electrons towards the dilution refrigerator. At the exit of the bending magnet, the electrons traverse through a 250-\si{\micro \meter}-thick beryllium exit window with a measured beam diameter of $\sim$3~\si{\milli \meter}.

After leaving the bending magnet, the electrons propagate 4.5~\si{\meter} in the accelerator hall and then through a 14.6~\si{\centi \meter} diameter penetration in a 1.8-\si{\meter}-thick reinforced concrete wall to the adjoining quantum hall. At the entrance of this penetration, multiple scattering in the atmosphere increases the beam's angular divergence to $\sim$77~\si{\milli \radian}~\cite{sgoudsmit1940-01, sgoudsmit1940-02}. The spatial extent of the beam has also expanded to an estimated 46~\si{\centi \meter} diameter at half maximum~\cite{bjmcparland1989, mhollmark2004}. The penetration between the accelerator hall and the quantum cell acts as a spatial collimator as electrons in the expanding beam are continuously lost to the walls. The electrons travel a final $\sim$1.3~\si{\meter} before reaching a 3.8~\si{\centi \meter} diameter titanium entrance window on the vacuum chamber of the dilution refrigerator. Once in the vacuum of this refrigerator, the electrons travel another 44.2~\si{\centi \meter} to the silicon die for a total distance of $\sim$8.1~\si{\meter} from the bending magnet exit.

The number of electrons incident on the silicon during an accelerator pulse must be statistically less than unity to investigate single electron effects on the device under test. The accelerator system was originally designed to produce 4~\si{\micro \second} wide pulses containing 1 to 350~\si{\nano \coulomb} of electron charge ($10^9 - 10^{12}$ electrons). To decrease the emitted charge, the thermionic electron gun's control system was modified to allow more precise adjustment of the grid potential. When electrons are not being injected into the accelerating structure, the grid potential is held at approximately -200~\si{\volt} below the cathode's potential of -18~\si{\kilo \volt}, thereby completely suppressing electron emission. For normal high charge injection, a positive 248~\si{\volt} pulse with a 4~\si{\micro \second} width is applied to the grid, raising it above the cathode potential with the current quickly reaching the space charge limit of $\sim$440~\si{\milli \ampere}. For low charge injection, the grid pulse is lowered to only 70~\si{\volt}, maintaining the grid below the cathode potential such that only a small area of the cathode emits. In this operating regime, the electron current is temperature limited and sensitive to changes in the cathode temperature.

\subsection{Dilution Refrigerator}

\begin{figure}[b]
\centering
\includegraphics[width = 0.98\textwidth]{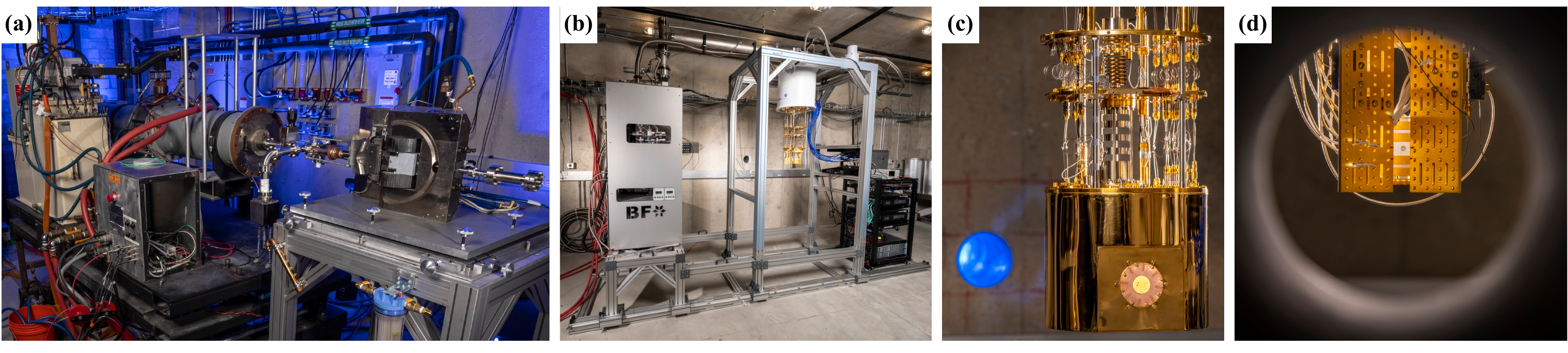}
\caption{
\textbf{CLIQUE Facility.} (a)~Image of the linear accelerator (left) and bending magnet (right). (b)~Image of the quantum hall, with dilution refrigerator and gas handling system mounted on a movable rail framework. (c)~Image of the wall penetration (illuminated with blue light) and modified mixing chamber stage shield (windows removed). (d)~Image of the qubit device packaging (gray, center) viewed from the accelerator hall through the wall penetration. 
}
\label{fig:supp_facility}
\end{figure}

The dilution refrigerator is a standard Bluefors LD400 system, shown in Fig.~\ref{fig:supp_facility}(b). The cryostat frame and gas handling system (GHS) have been raised so that the sample space below the mixing chamber is aligned with the linac beam line, $\sim$1.5~\si{\meter} off the ground. The supports for the refrigerator are mounted to a linear rail system capable of $\sim$1~\si{\meter} of horizontal motion while operating at base temperature. This degree of freedom is used to fine-tune the alignment with the wall penetration and to confirm that stray ionizing or RF radiation from linac operation does not affect the qubits when pushed out of the beam line. Added structural supports between the cryostat frame and GHS shown in Fig.~\ref{fig:supp_facility}(b) are removed during qubit measurements to minimize vibrational coupling of the two components. 

The modified mixing chamber shield is shown in Fig.~\ref{fig:supp_facility}(c). These are optical access shields provided by Bluefors with 3-\si{\centi \meter} diameter windows at each shield layer facing both the wall penetration and 180\si{\degree} from the wall. The interior shield windows facing the incident electron beam have 25-\si{\micro \meter}-thick contaminant-free high-vacuum aluminum foil covering the aperture. The exterior vacuum shield has an additional $\sim$20-\si{\centi \meter}-long nipple flange fitting with a 3~\si{\centi \meter} diameter, 25-\si{\micro \meter}-thick light-tight titanium film creating the vacuum seal. The reverse side windows have thick, $\sim$0.5~\si{\milli \meter} aluminum or copper blanks covering the apertures. These thinned thermal shields do not cause a measurable increase in the base mixing chamber plate temperature ($\sim$10$\si{\milli \kelvin}$). 

The electron path within the mixing chamber is shown in Fig.~\ref{fig:supp_facility}(d). The qubit package is centrally mounted behind two RF component mounting brackets. The backside of the package is aluminum and faces the incident electron beam. It is thinned in a 5~\si{\milli \meter} diameter area to $\sim$800~\si{\micro \meter} directly behind the qubit chip, which sits atop four narrow posts at each corner. Not shown is a \si{\milli \meter}-thick cryo-NETIC mu-metal shield that surrounds the qubit and TWPA packages. In total, there is 25~\si{\micro \meter} of titanium, $\sim$900~\si{\micro \meter} of aluminum, and 1~\si{\milli \meter} of mu-metal (mostly nickel-iron alloy) in the electron path before reaching the silicon qubit chip.

\subsection{Scintillator Detector}

\begin{figure}
\begin{center}
\includegraphics[width = 0.49\textwidth]{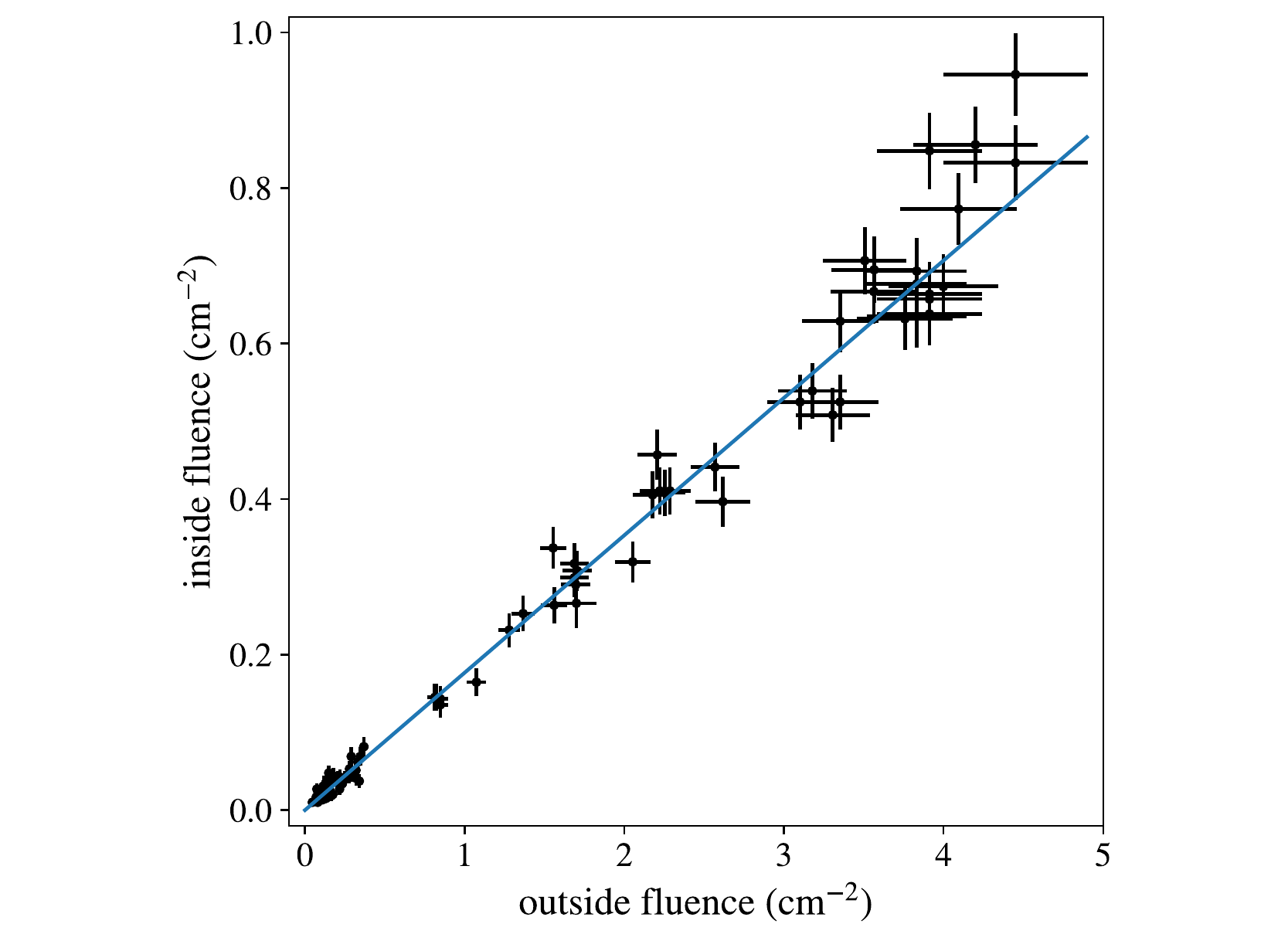}
\caption{\textbf{Scintillator Fluence Calibration.}
Measured electron fluence at the location of the silicon die versus the 
electron fluence outside the dilution refrigerator. A linear fit is overlaid 
on the data and shows a ratio of 0.177 between the inside and outside fluences.
}
\label{fig:enum2venum1}
\end{center}
\end{figure}

The particle fluence at the entrance to the dilution refrigerator was continuously monitored by a $4$-\si{\milli \meter}-thick plastic scintillator with a 1 by 1~\si{\centi \meter} cross section, mounted to a photomultiplier tube. This detector was 5~\si{\centi \meter} below the center of the dilution refrigerator's entrance port and 7.4~\si{\meter} from the exit of the bending magnet. The detector signal was amplified with a full width half maximum Gaussian shaping time of 4.9~\si{\micro \second} that integrated the charge collected in a single accelerator pulse. The peak height of this amplified signal was digitized by an analog to digital converter in a multiparameter list-mode data acquisition system. In addition, the arrival time of accelerator triggers and the detector signals are recorded, allowing coincidences to be identified.

Figure~\ref{fig:intro}(b) in the main text shows averaged energy deposition spectra in the plastic scintillator at two different biases applied to the electron gun's grid. At the lower grid bias, 6.7\si{\percent} of the accelerator pulses had a coincident radiation signal, ensuring that no more than one electron strikes the scintillator. In this spectrum, there is one peak at 634~\si{\kilo \electronvolt}, corresponding to the most probable energy deposition for a single electron passing through the plastic scintillator. Raising the grid bias increased the detection probability to 74.9\si{\percent} of the accelerator pulses and peaks at 634, 1268 and 1902~\si{\kilo \electronvolt} are visible in the spectrum, corresponding to one, two and three electrons traversing the scintillator in a single accelerator pulse. While not shown, continuing to raise the grid bias eventually increased the detection probability to unity (i.e. a radiation event is detected every accelerator pulse) and peaks at integer multiples of 634~\si{\kilo \electronvolt} cannot be discerned in the spectrum.

The fluence on the scintillator can be determined by considering the number of incident particles (electrons and photons) during an accelerator pulse follows a Poisson distribution. Since the detector's signal is integrated over the accelerator pulse, it can only detect or not detect a single radiation event in this period. The overall probability to detect is then the compound distribution of the Poisson distribution with probability to detect one or more individual particles, $1-\left(1-\epsilon_{d}\right)^{n_{p}}$, resulting in 
\begin{equation}
P_{d} = 1-e^{-\lambda_{p}\cdot \epsilon_{d}}, 
\label{equ:elecproddet} 
\end{equation}
where $\lambda_{p}$ is the mean number of electrons and $\epsilon_{d}$ is the intrinsic efficiency of the detector for a single particle. The mean number of particles can then be found
\begin{equation}
\lambda_{p} = \frac{1}{\epsilon_{d}} \ln\left( \frac{1}{1-P_{d}} \right),
\label{equ:elecnumb1}
\end{equation}
with $P_{d}$ being nothing more than the fraction of accelerator pulses in which a radiation event was detected; an easily measurable quantity. The detection efficiency for electrons is close to unity, while the photon detection is $\sim$1\si{\percent}. Hence, $\lambda_{p}$ is dominated by electrons and is a good proxy for the electron fluence.

To correlate the electron fluence measured outside the dilution refrigerator to the electron fluence incident on the silicon die, a plastic scintillator was mounted inside the refrigerator when it was at room temperature. This second electron detector was identical to the outside detector but had an $800$-\si{\micro \meter}-thick aluminum plate in front of the scintillator to mimic the qubit packaging.  Figure~\ref{fig:enum2venum1} shows the measured linear correlation between the electron fluence at the location of the silicon die versus the electron fluence outside the dilution refrigerator. The ratio of the fluences measured inside versus outside the dilution refrigerator was $R_{io}=0.177\pm 0.002$.

\subsection{MCNP Modeling}

\begin{figure}
\begin{center}
\includegraphics[width=5.5in]{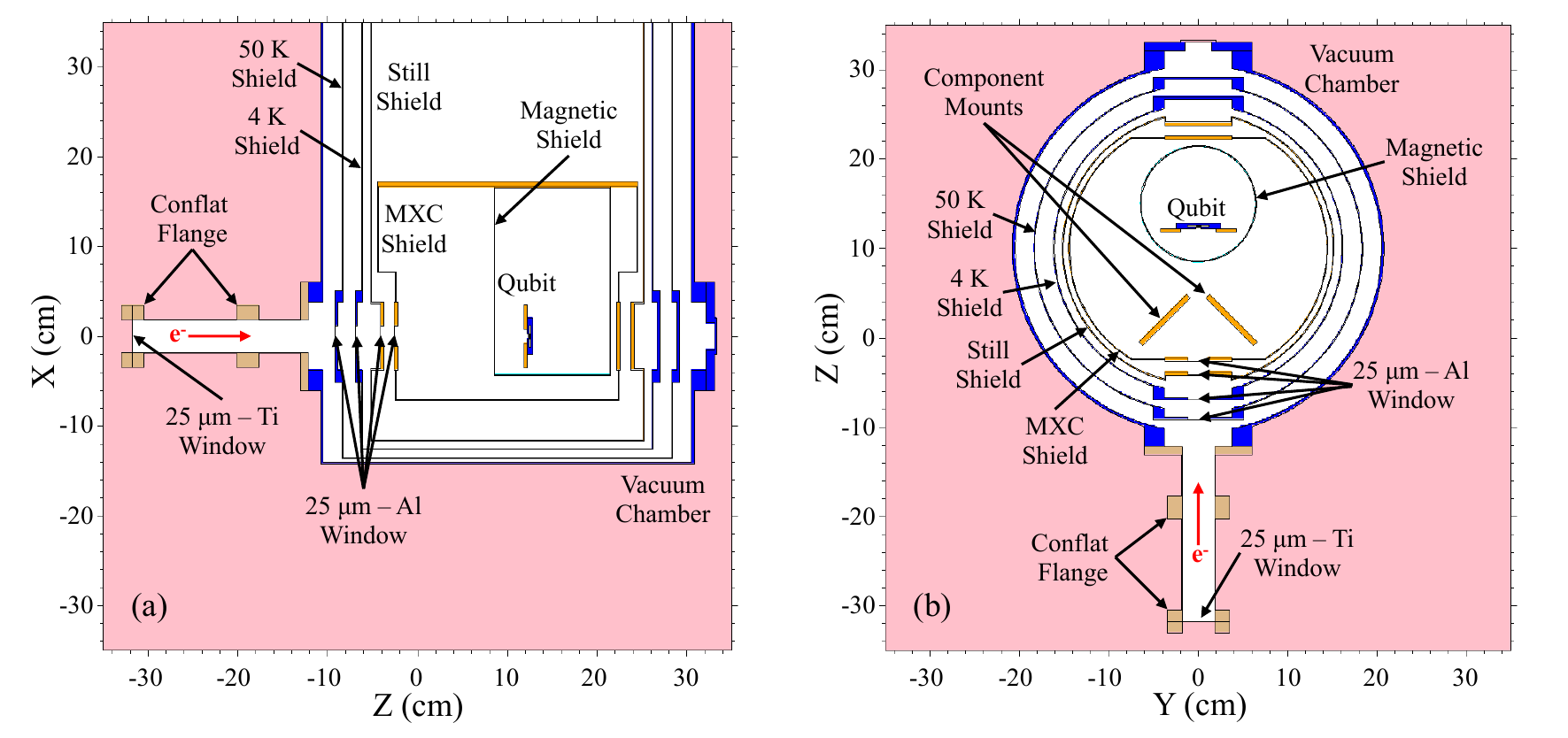}
\caption{\textbf{Physical Model Details.}
Cross sectional schematic of the dilution refrigerator model used in the radiation transport simulations as viewed from the top-down (left) and the side (right). The colors represent the different materials in the model; atmosphere (pink), stainless steel (light brown), aluminum (blue), copper (orange) and no material (white). The electron beam propagation direction is along the $z$-axis.
}
\label{fig:mcnp_geom}
\end{center}
\end{figure}

The Monte Carlo N-Particle (MCNP) software was used to model the transport of the electrons from the accelerator to the device under test and to track the energy deposition in the silicon~\cite{kulesza2024}. These simulations started with a three dimensional representation (Fig.~\ref{fig:mcnp_geom}) of the objects and materials in the laboratory that were important to the electron transport, including the bending magnet's 250-\si{\micro \meter}-thick beryllium exit window, the 1.8-\si{\meter}-thick concrete wall with the 14.6~\si{\centi \meter} diameter penetration and the components of the dilution refrigerator. The space between the accelerator exit and refrigerator entrance was filled with atmosphere. The refrigerator model includes the outer aluminum vacuum chamber with the 25-\si{\micro \meter}-thick titanium window and associated vacuum hardware, the thermal shields with their 25-\si{\micro \meter}-thick aluminum windows, two 6-\si{\milli \meter}-thick copper component mounts, the 1-\si{\milli \meter}-thick mumetal magnetic shield, the quantum device package and support, and the 350-\si{\micro \meter}-thick slicon die with a 5 by 5~\si{\milli \meter} cross section.

In the simulations, 18.5~\si{\mega \electronvolt} monoenergetic electrons were launched from the exit side of the bending magnet's beryllium window within an initial 3~\si{\milli \meter} diameter circular area. As in the experiments, the simulated electrons then propagated through the model's accelerator hall, concrete wall penetration, and the quantum hall. During this transport, the full electromagnetic shower was simulated with secondary particle creation (photons, electrons and positrons), elastic scattering and particle energy loss. All particles were tracked down to 10~\si{\kilo \electronvolt}. The EL03 electron and MCPLIB04 photoatomic data libraries were utilized and contained all the needed cross section and stopping-power data~\cite{adams2000,white2003}. 

The simulated electron and photon fluences were monitored at the entrance and exit of the penetration in the concrete wall, at the plastic scintillator outside the refrigerator, and at the front surface of the silicon chip. At the entrance of penetration in the concrete wall the simulated beam's most probable electron energy was 17.6~\si{\mega \electronvolt} and its full width at half maximum had grown to $\sim$45~\si{\centi \meter}, both of which are in good agreement with the analytic estimates~\cite{landu1965, hbischsel1988, mjberger2005,bjmcparland1989, mhollmark2004}. At the penetration's exit, the most probable energy had decreased to 17.1~\si{\mega \electronvolt}, the beam's spatial extent was limited to the 14.6~\si{\centi \meter} diameter and the fluence decreased by a factor of $\sim$5 owing to the beam's divergence through the wall penetration length. As the beam traveled the 1.1~\si{\meter} from this exit to the plastic scintillator, the beam again spatially expanded with the fluence decreasing an additional factor of $\sim$2. The most probable electron energy incident on the scintillator was 16.7~\si{\mega \electronvolt}. The electrons then traversed the final 44.2~\si{\centi \meter} in the vacuum of the refrigerator from the titanium entrance window to the silicon die, with its 1~\si{\milli \meter} mu-metal magnetic shield being the largest source of scattering and energy loss. At the surface of the silicon the most probable electron energy was 15~\si{\mega \electronvolt}, and fluence had decreased by another factor of 5.2. Hence the ratio of the simulated fluence at the silicon die versus the plastic scintillator outside the refrigerator was 0.19, which is consistent with the measured value of $R_{io}=0.177\pm 0.002$ obtained using two plastic scintillators (see Fig.~\ref{fig:enum2venum1}).

\begin{figure}
\begin{center}
\includegraphics[width=0.49\textwidth]{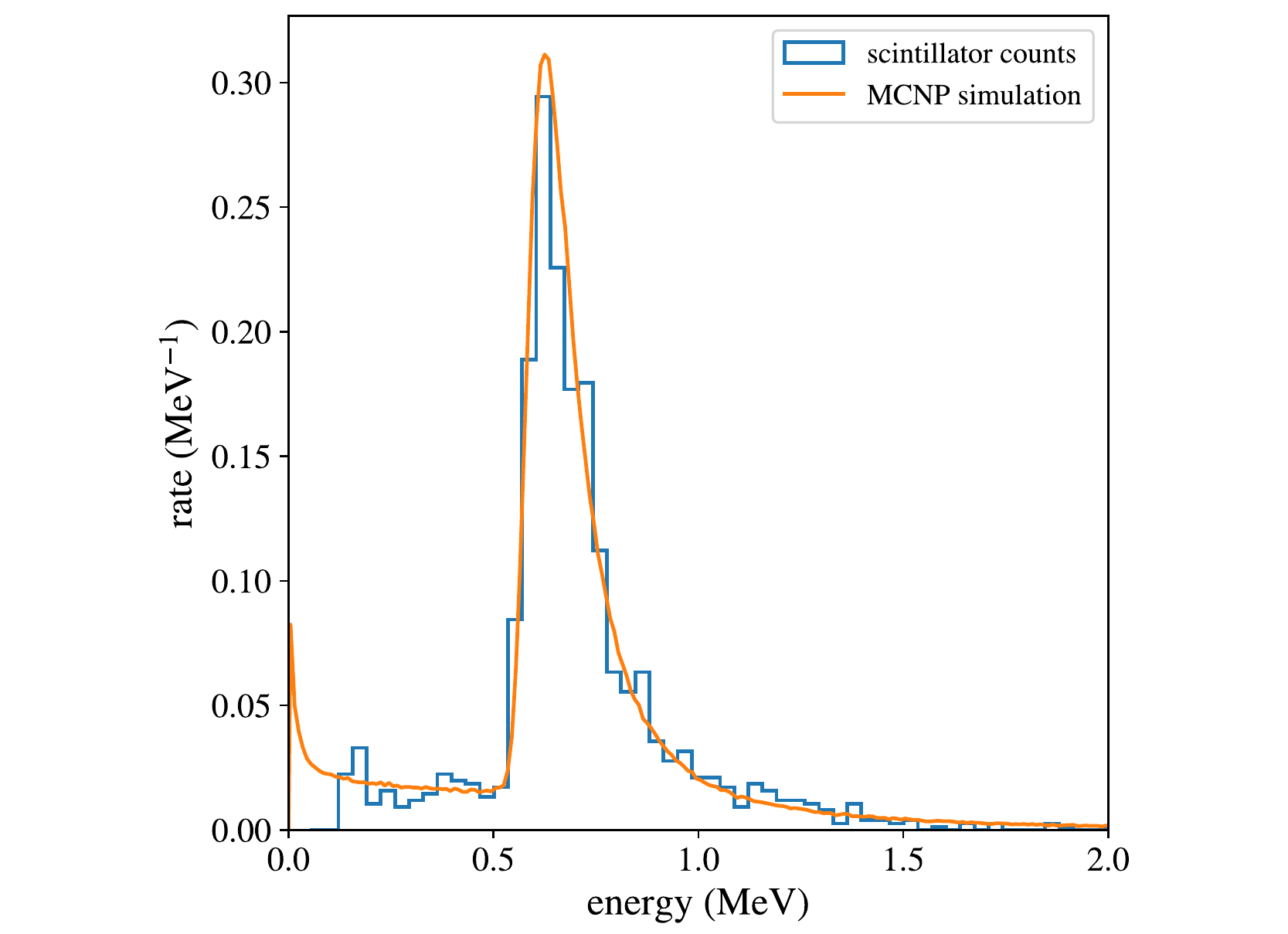}
\caption{\textbf{Modeling Comparison.} Measured energy deposition distribution in the scintillator (blue) compared to the MNCP simulation results (orange). The linac electron fluence was tuned to result in a low $\sim$7\si{\percent} of pulses with a coincident radiation signal. The simulated spectrum was normalized to the observed peak, showing good agreement in the distribution.
}
\label{fig:edep_simvmeas}
\end{center}
\end{figure}

The energy deposited in the silicon and the plastic scintillator was also tallied for every electron launched from the exit window of the accelerator's bending magnet. These energy tallies included deposition from all secondary particles. Figure~\ref{fig:edep_simvmeas} compares simulated and measured energy deposition in the plastic scintillator. The measured spectrum is taken from Fig.~\ref{fig:intro}(b) in the main text, when the grid bias was pulsed by 65~\si{\volt}, resulting in 6.7\si{\percent} of the accelerator pulses having a coincident radiation signal. The simulated spectrum was normalized to give this same 6.7\si{\percent} rate giving good agreement between the measured and simulated spectra. Figure~\ref{fig:intro}(c) in the main text shows the simulated energy deposition in the silicon. The most probable energy deposition is 100~\si{\kilo \electronvolt} and is consistent with analytic estimates~\cite{landu1965, hbischsel1988, mjberger2005}. Overlaid on Fig.~\ref{fig:intro}(c) in the main text is the energy deposition from negative muons demonstrating the similarity in energy deposition with electrons. In these simulations, monoenergetic 400~\si{\mega \electronvolt} muons were launched from the exit of the accelerator's bending magnet and then propagated to the silicon die through the same laboratory and refrigerator model as was used in the electron transport simulations. The initial 400~\si{\mega \electronvolt} muon energy was chosen because it is the most probable cosmogenic muon energy at sea level~\cite{rastin1984}.

\section{Quantum Control}

\begin{figure}
\centering
\includegraphics[width = 0.98\textwidth]{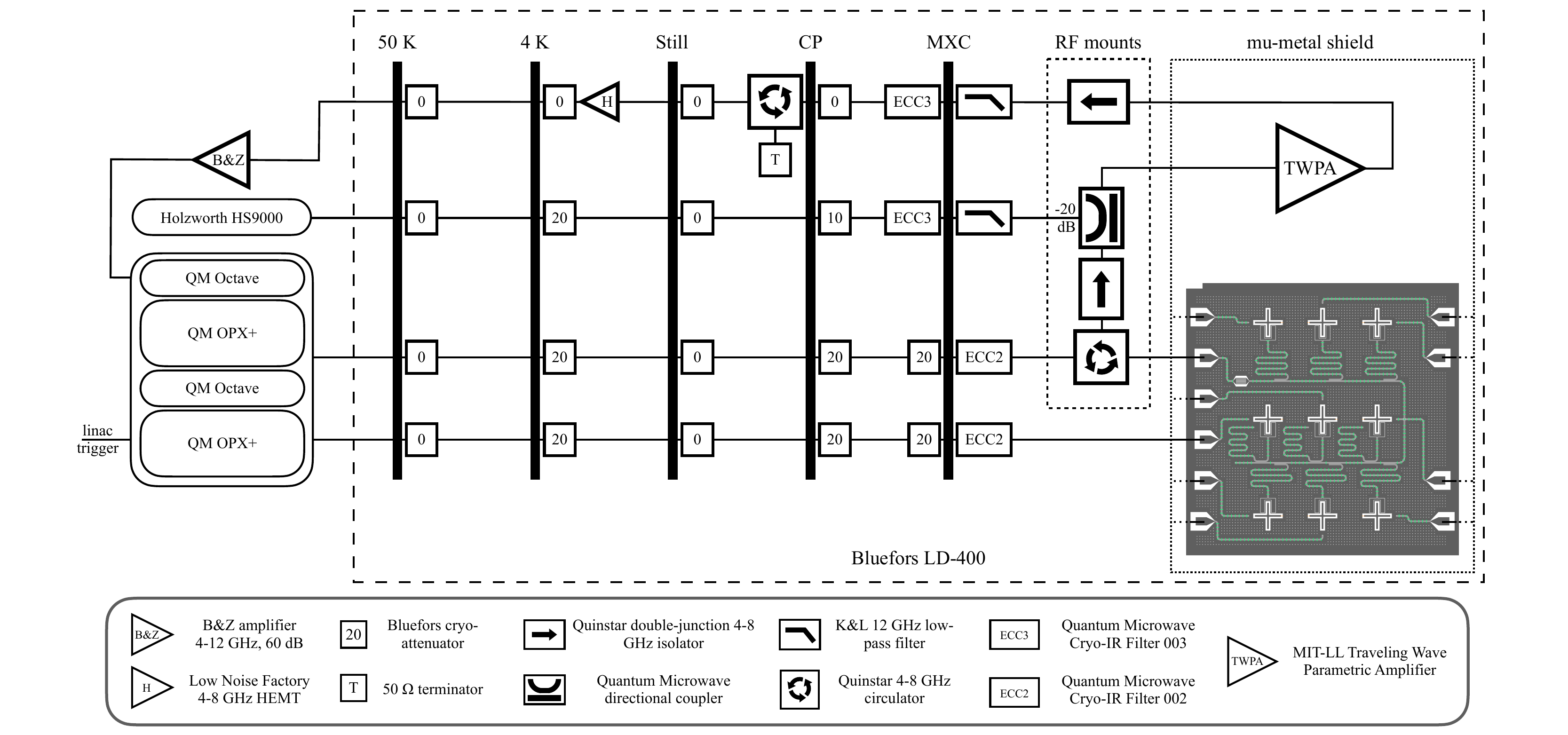}
\caption{
\textbf{Wiring Diagram.} Only one qubit control line is shown, but all nine qubits have individual control.
}
\label{fig:supp_wiring}
\end{figure}

\subsection{Fridge Wiring}

The attenuation, filtering, RF components and amplifiers used in this work are shown in Fig.~\ref{fig:supp_wiring}. Each qubit has individual voltage control sourced from a set of Quantum Machines OPX+ and Octave equipment. The measurement tones on the single shared readout line are amplified at various stages and integrated by the Quantum Machines stack, which additionally records the linac trigger alongside the qubit measurements. A Holzworth RF synthesizer provides the pump tone for the MIT Lincoln Laboratory supplied traveling wave parametric amplifier (TWPA). 

\subsection{Qubit Details}

The 5 by 5~\si{\milli \meter} transmon qubit chip (see Fig.~\ref{fig:supp_wiring}) was designed by the authors and fabricated at MIT Lincoln Laboratory. All device layers are aluminum deposited on a 350-\si{\micro \meter}-thick silicon substrate. The junction orientation, qubit frequencies, readout frequencies, and average $T_1$ times for the eight qubits used in this work are listed in Table~\ref{tab:qubitparams}. The unused ninth qubit (q32) was excluded due to poor performance.

\begin{table}
\caption{\label{tab:qubitparams}\textbf{Qubit Parameters.} Index keys for qubits are shown in Fig.~\ref{fig:intro}(e) of the main text. Frequencies reported are the $\ket{0}$ to $\ket{1}$ state transitions of the qubits ($f_q$) and fundamental frequencies of the readout resonators ($f_{RO}$). Reported $T_1$ values are the weighted mean of estimated fit values for the explicit $T_1$ experiments throughout the relaxation and excitation data collection day. Reported uncertainties are propagated errors of the standard deviation of the individual fits.}
\begin{ruledtabular}
\begin{tabular}{ccccc}
index & junction orientation & $f_q$ (\si{\giga \hertz}) & $f_{RO}$ (\si{\giga \hertz}) & $T_1$ (\si{\micro \second})\\
\hline
q11 & low-gap-island & 4.197 & 6.619 & 35$\pm$7 \\
q12 & high-gap-island & 4.377 & 6.683 & 72$\pm$13 \\
q13 & high-gap-island & 3.964 & 6.748 & 36$\pm$5 \\
q21 & low-gap-island & 4.268 & 6.839 & 42$\pm$8 \\
q22 & low-gap-island & 3.852 & 6.901 & 61$\pm$7 \\
q23 & high-gap-island & 4.089 & 6.963 & 73$\pm$19 \\
q31 & low-gap-island & 4.029 & 6.972 & 42$\pm$7 \\
q33 & high-gap-island & 4.352 & 7.118 & 80$\pm$17 \\
\end{tabular}
\end{ruledtabular}
\end{table}

\subsection{Control Schemes}

The pulse length for a full $\pi$ rotation to prepare the $\ket{1}$ or $\ket{0}$ state in all qubits was 80~\si{\nano \second}. The pulse lengths for $\sqrt{x}$ and $\sqrt{y}$ gates were 40~\si{\nano \second}. The total readout times were 4~\si{\micro \second}, with various splits between the measurement pulse and ring-down time depending on the qubit-specific resonator line-widths. The reset protocol used the result of the previous shot's measurement to conditionally apply a $\pi$ pulse when in the $\ket{1}$ state, resulting in reliable $\ket{0}$ state preparation $>$90~\si{\percent}. Three-state discrimination of idle control and repeated measurement of q22 estimated a qubit temperature below 50~\si{\milli \kelvin}. 

For the detection of relaxation (excitation) errors, qubits were reset, prepared in the $\ket{1}$ ($\ket{0}$) state with a $\pi$ (identity) pulse, left to evolve for the length of the detection delay, then measured. For the detection of detuning errors, the set qubit frequencies for the control hardware were detuned by +100~\si{\kilo \hertz}, then the qubits were reset, prepared in a superposition with a $\sqrt{x}$ pulse, left to evolve for a varying detection delay, projected back to the computational basis with a $\sqrt{y}$ pulse, and measured. The detection delay cycled between 10 different times ranging from 32~\si{\nano \second} to 9~\si{\micro \second}. For both experimental schemes, the state ($\ket{0}$ or $\ket{1}$) of each qubit was recorded along with the presence or absence of the linac trigger for each shot. 

\section{Data and Analysis}

\subsection{Data Collection}

All linac-correlated relaxation and excitation data were collected in the span of a single day. The linac trigger rate was 10~\si{\hertz}, and the current was tuned so that $\sim$20\si{\percent} of the triggers resulted in an electron collision with the qubit chip. The qubits were re-tuned every 15 minutes in between sets of measurements. Each set consisted of 14 different measurement runs with varying qubit prep ($\ket{0}$ or $\ket{1}$) and detection delay (0.2, 0.6, 1, 2, 4, 8, or 12~\si{\micro \second}), each of which contained $5 \times 10^6$ individual measurement shots. In total, $1.23 \times 10^9$ measurements were performed, capturing 97,733 linac triggers. 

For the Ramsey detuning measurements, all data were collected in a (separate) single day. The linac conditions were the same and the qubits were re-tuned every 5~\si{\minute} as unintentional changes to the effective detuning would more sensitively affect the measurement. The Ramsey measurement, which consisted of 0.032, 1, 2, 3, 4, 5, 6, 7, 8, and 9~\si{\micro \second} detection delays, contained $10^7$ measurement shots and was repeated three times between each re-tuning. This routine was performed eight times, for a total of $2.4 \times 10^8$ measurement shots, capturing 21,694 linac triggers. 

\subsection{Event Classification}

\begin{figure}
\centering
\includegraphics[width = 0.98\textwidth]{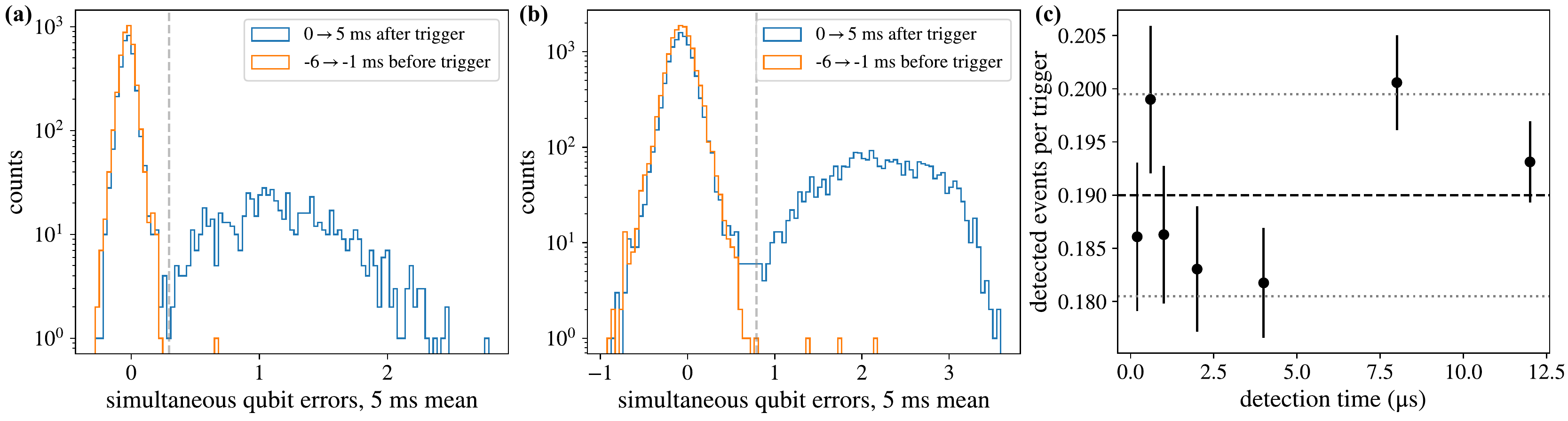}
\caption{
\textbf{Relaxation Event Detection.} Threshold determination (dashed gray line) of eventful triggers for (a)~0.2~\si{\micro \second} and (b)~12~\si{\micro \second} detection time. (c)~Detected event rate for relaxation detection across all seven detection delays. Error bars are the $\sqrt{N}$ standard error of a count. Black dashed line is the arithmetic mean, dotted gray lines are $\pm$5\si{\percent} off the mean.
}
\label{fig:supp_events}
\end{figure}

For relaxation detection, correlated qubit error events were distinguished from background errors using the time trace sum of all eight qubit states. For each trigger, the mean of the 5~\si{\milli \second} following the trigger and the mean -6 to -1~\si{\milli \second} before the trigger were calculated. These two sets of mean values were binned into histograms as shown in Fig.~\ref{fig:supp_events}(a-b) to determine a threshold for ``eventful" triggers. This threshold varied between detection delays. Figure~\ref{fig:supp_events}(c) shows the event rate (fraction of triggers with identified events) across all detection delays, which demonstrates the stability of the system and reliability of the event detection scheme. Though more complex event extraction techniques could be implemented (\textit{e.g.} matched filters, binomial classification checks), this analysis was sufficient for our purposes.

For excitation detection, because the error signatures are much smaller with a more complex shape (see Fig.~\ref{fig:prep10} in the main text), the event statistics of relaxation detection were used to create an eventful trigger threshold. The excitation detection data was similarly organized by calculating the mean of the four high-gap-island qubits across the first ms after each linac trigger. The threshold for eventful triggers was set by assuming an electron collision rate identical to the average event rate found in relaxation detection [mean of Fig.~\ref{fig:supp_events}(c)]. Excitation and relaxation data runs were interspersed, so the expectation of a constant event rate between data sets should be valid.

\subsection{Rate Fit}

As described in the main text, a phenomenological two-level-system model was used to fit the relaxation- and excitation-sensitive experiments across all detection delays to a single set of rate equations. The primary goal was to extract more accurate decay times from the $\ket{0}$ ($\ket{1}$) state preparation data as the presence of relaxation (excitation) will affect the visibility of excitation (relaxation) errors. The number of fit and fixed parameters was kept to a minimum using auxiliary information within the data (baseline errors) and additional measurements ($T_1$ estimates). 

\begin{figure}
\centering
\includegraphics[width = 0.49\textwidth]{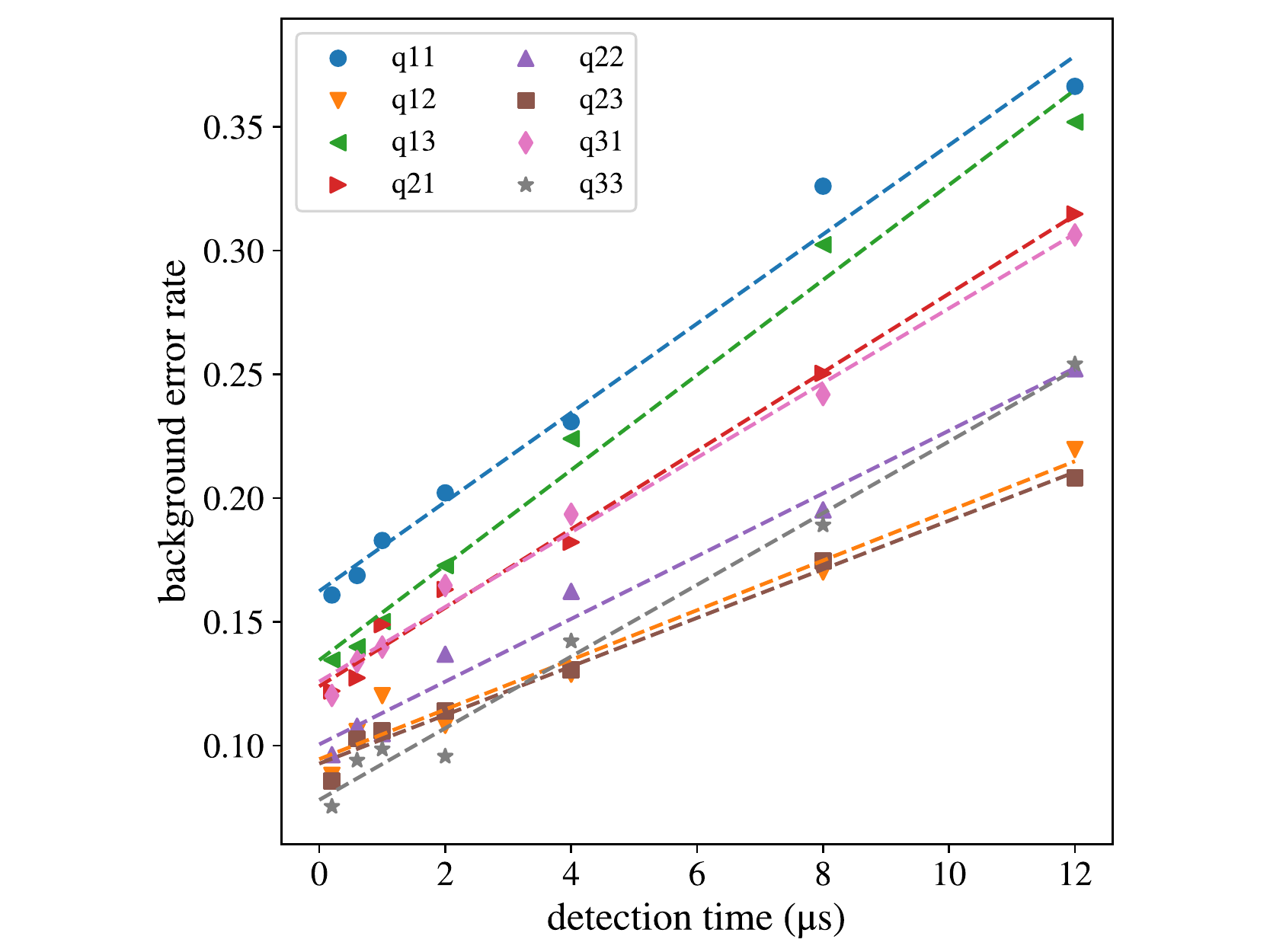}
\caption{
\textbf{SPAM Fidelity Estimate.} Estimates of the $\ket{1}$ state preparation and measurement (SPAM) error are complicated by non-zero detection time between preparation and measurement. To overcome this, the relaxation rates measured for each qubit across all detection times are fit to lines (dashed lines) to extract the detection time~=~0 limit. These estimates are higher than typical for these qubits due to the active reset protocol and simultaneous operation.
}
\label{fig:supp_back_rate}
\end{figure}

\begin{figure}
\centering
\includegraphics[width = 0.98\textwidth]{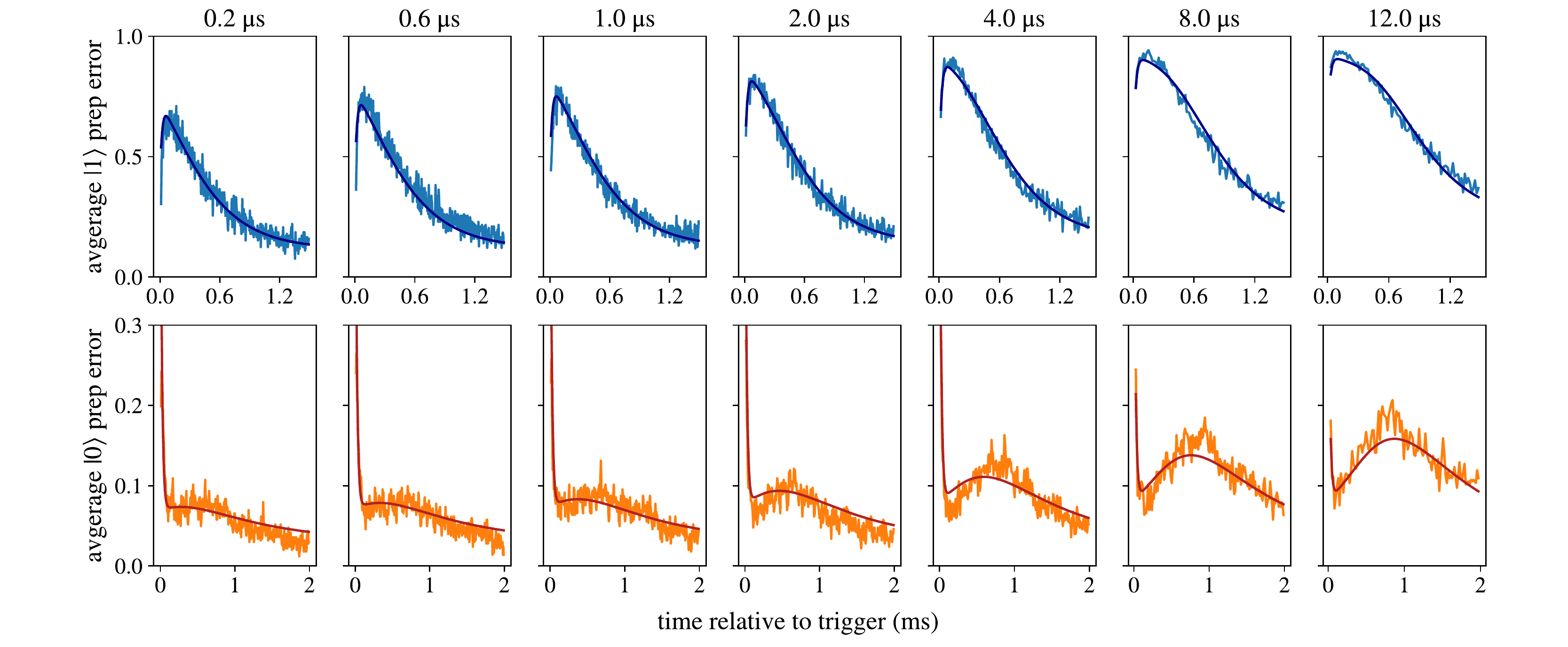}
\caption{
\textbf{Complete Fit Results for q23.} Data (blue / orange) and fits (dark blue / dark red) for $\ket{1}$ state preparation (top) and $\ket{0}$ state preparation (bottom) across all seven detection times (columns) for q23. All 14 data sets are fit to the same six parameter model that captures competing relaxation and excitation errors, the relaxation error saturation at long detection times, and multiple decay times.
}
\label{fig:supp_fits}
\end{figure}

\begin{table}
\caption{\label{tab:suppfits}\textbf{Fit parameter results for all qubits}. Reported uncertainties are standard deviations from the fits.}
\begin{ruledtabular}
\begin{tabular}{cccccccc}
qubit & junction orientation & $A_{rel}$ (\si{\kilo \hertz}) & $\tau_{rel}$ (\si{\micro \second}) & $A_{exc_1}$ (\si{\kilo \hertz}) & $\tau_{exc_1}$ (\si{\micro \second}) & $A_{exc_2}$ (\si{\kilo \hertz}) & $\tau_{exc_2}$ (\si{\micro \second})\\
q11 & low-gap-island & 294$\pm$1 & 8080$\pm$10 & 183$\pm$20 & 20$\pm$2 & N/A & N/A \\
q12 & high-gap-island & 527$\pm$4 & 388$\pm$2 & 38$\pm$1 & 1030$\pm$30 & 480$\pm$50 & 15$\pm$1 \\
q13 & high-gap-island & 383$\pm$4 & 283$\pm$2 & 45$\pm$1 & 1010$\pm$21 & 360$\pm$30 & 19$\pm$1 \\
q21 & low-gap-island & 467$\pm$1 & 9110$\pm$11 & 190$\pm$10 & 34$\pm$2 & N/A & N/A \\
q22 & low-gap-island & 272$\pm$1 & 6240$\pm$10 & 162$\pm$8 & 42$\pm$2 & N/A & N/A \\
q23 & high-gap-island & 611$\pm$5 & 370$\pm$2 & 46$\pm$1 & 960$\pm$20 & 760$\pm$50 & 17$\pm$1 \\
q31 & low-gap-island & 273$\pm$1 & 7880$\pm$10 & 137$\pm$12 & 29$\pm$2 & N/A & N/A \\
q33 & high-gap-island & 487$\pm$3 & 414$\pm$2 & 34$\pm$1 & 1000$\pm$30 & 620$\pm$60 & 13$\pm$1 \\
\end{tabular}
\end{ruledtabular}
\end{table}

The state vector $P(t)$ in Eq.~\ref{eq:Pt} is evolved over the time of a single shot (detection delay plus half the measurement time) using \texttt{scipy.integrate.odeint}~\cite{2020SciPy-NMeth}. A correction matrix $M_{corr}$ is used to to capture the state preparation and measurement (SPAM) infidelity. Each shot is assumed to be prepared in $M_{corr} \cdot [1, 0]$ or $M_{corr} \cdot [0, 1]$ and after evolution is again multiplied by $M_{corr}$ (the infidelity of an $X$ gate is assumed to be much less than $M_{corr}$). The correction matrix is:
\begin{equation}
M_{corr} = \begin{pmatrix}
1-\frac{1}{2}\epsilon_0 && \frac{1}{2}\epsilon_1 \\
\frac{1}{2}\epsilon_0 && 1-\frac{1}{2}\epsilon_1
\end{pmatrix}
\end{equation}
where $\epsilon_0$ is the approximate $\ket{0}$ SPAM infidelity and $\epsilon_1$ is the approximate $\ket{1}$ SPAM infidelity. The equivalence of preparation and measurement infidelity is considered valid due to the active reset protocol resulting in the state preparation's dependence on the previous shot's measurement. The factor of $\frac{1}{2}$ accounts for the correction matrix being applied at the beginning and end of the shot evolution. The averaged error rate across the 1000 points preceding the linac trigger for $\ket{0}$ state preparation is used as the baseline $\epsilon_0$ value, unique to each qubit and detection time. Non-zero detection time will increase the ``background" relaxation errors and affect the baseline error rate of $\ket{1}$ state preparation. For this reason, the averaged error rate across the 1000 points preceding the linac trigger for $\ket{1}$ state preparation across all detection times is fit to a line for each qubit and the extrapolated value at zero detection time is used for $\epsilon_1$ across all detection times (see Fig.~\ref{fig:supp_back_rate}).

For each qubit with the high(low)-gap-island junction orientation, both state preparations and all seven detection delays are fit simultaneously to estimate the six (four) fit parameters. Different fit lengths are used for $\ket{0}$ and $\ket{1}$ state preparation to minimize overemphasis on regions with minimal information ($\ket{1}$ state preparation returns to baseline quicker). Figure~\ref{fig:supp_fits} shows the fit results for q23 across all seven detection delays, showing good agreement for the same fit parameters across all data. Table~\ref{tab:suppfits} lists the extracted fit parameters for all qubits.

\subsection{Ramsey Data Processing}

\begin{figure}
\centering
\includegraphics[width = 0.98\textwidth]{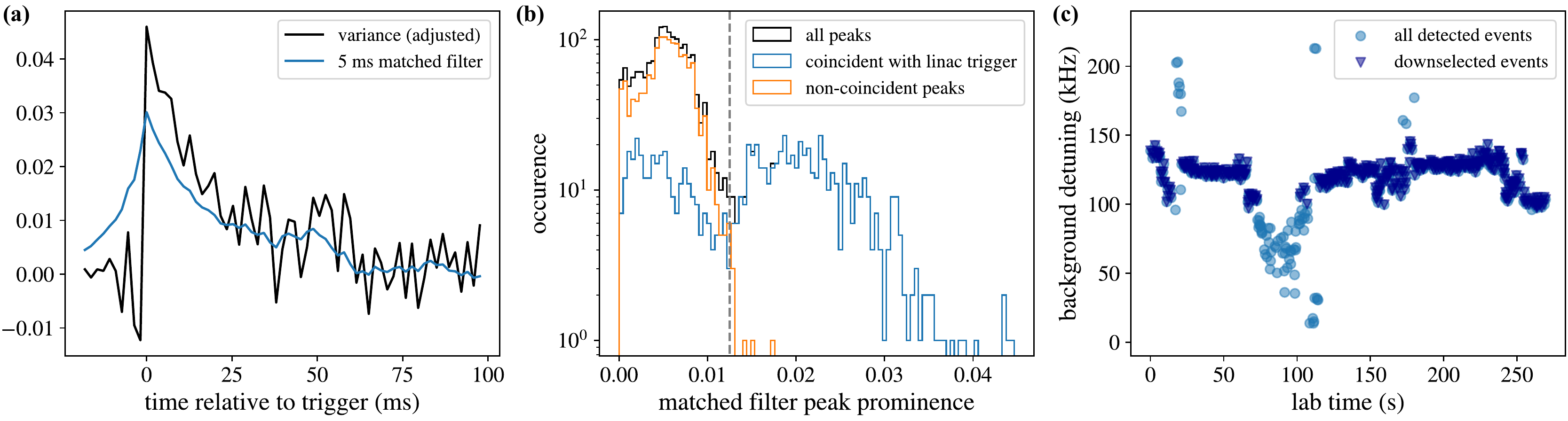}
\caption{
\textbf{Ramsey Event Down Selection.} (a)~Example linac-induced detuning/dephasing event. As described in the text, the variance of 20 Ramsey sequences was calculated, which down-sampled the data into 1.8~\si{\milli \second} steps. Plotted in black is the variance, inverted and offset removed so that events appear as peaks deviating from zero. In blue, a matched filter with 5~\si{\milli \second} decay time shows a peak at the start of the event. (b)~Example event classification for one set of three data runs (2711 linac triggers). The peaks and prominences in the matched filter data were identified using \texttt{scipy.optimize}. The prominences are compared and divided into those coincident with the linac trigger where the peak is found less than six points from the trigger (blue) and those non-coincident (orange). As shown, there is an obvious threshold line to divide real events from background noise. For this data, there were 542 detected events (20\si{\percent} of triggers). (c)~Identified events are down selected to exclude regions of the data collection when the qubit had quasi-statically detuned from the desired offset ($\sim$100~\si{\kilo \hertz}). The extracted Ramsey oscillation frequency from the 500 sequences preceding each trigger is plotted in blue circles. The average (excluding strong outliers) is calculated and those that fall within 20\si{\percent} of the mean is plotted in dark blue triangles. Now only 459 in total, these events were then aligned and averaged together as shown in Fig.~\ref{fig:detuning}(a) in the main text.}
\label{fig:supp_detuning_back}
\end{figure}

Event classification and processing for the Ramsey-like sequence differed significantly from the excitation and relaxation detection due to the oscillatory nature of the qubit response across detection time. The responses of all eight qubits were divided into 200 point segments (20 full Ramsey sequences, $\sim$1.8~\si{\milli \second}) and the variance calculated. For segments with long $T_1$ and $T_2$ times, the variance should be high, as the Ramsey oscillations should be large. When poisoned by quasiparticles, these lifetimes drop and the variance will decrease. This provides a method to convert the oscillating raw signals of eight qubits into an easily understandable single stream of data that dips during transient increases in QP density. A matched filter with a 5~\si{\milli \second} decay was then applied and a prominence threshold used to identify events, as shown in Fig.~\ref{fig:supp_detuning_back}(a-b).

As the device was routinely re-tuned between sets of three $10^7$ measurement shots (affecting the intentional $\sim$100~\si{\kilo \hertz} detuning), the following processing steps were performed separately for each qubit and each set of three runs. The identified events were further down-selected to exclude data where the qubit frequency had significantly shifted independent of the linac triggers. The 500 Ramsey sequences ($\sim$22.5~\si{\milli \second}) preceding each identified event were averaged and fit to a sinusoid to extract the baseline detuning. With all extracted baselines across the $\sim$5~\si{\minute} set of runs, the average qubit detuning was calculated and only individual events $<$20\si{\percent} deviation from the mean were kept. See Fig.~\ref{fig:supp_detuning_back}(c) for an example.

These down-selected events were then aligned to one another with respect to the trigger, i.e., if the linac trigger arrived in the $n^{th}$ detection time the data would be shifted by $n$ steps (up to 10, the number of detection times). This reduces the timing accuracy from a single shot to the total time it takes for all 10 Ramsey steps ($\sim$90~\si{\micro \second}). All aligned events are averaged together and further down-sampled by a factor of 2 (reducing the timing to $\sim$180~\si{\micro \second}). These highly averaged curves are then fit to decaying sinusoids to extract frequency shifts and drops in the envelope decay: see Fig.~\ref{fig:detuning}(a) in the main text.

This process of event down-selection and fitting is repeated for each qubit and 3-round measurement set. The intentional detuning in each measurement set is subtracted out and the fit results for all measurement sets are averaged together for each qubit. This yields the final result of excess negative detuning and dephasing (envelope decay) shown in Fig.~\ref{fig:detuning}(b-c) in the main text. The error bars shown are the propagated errors of the standard deviations from the decaying sinusoid fits.

\subsection{Control Tests}

\begin{figure}
\centering
\includegraphics[width = 0.98\textwidth]{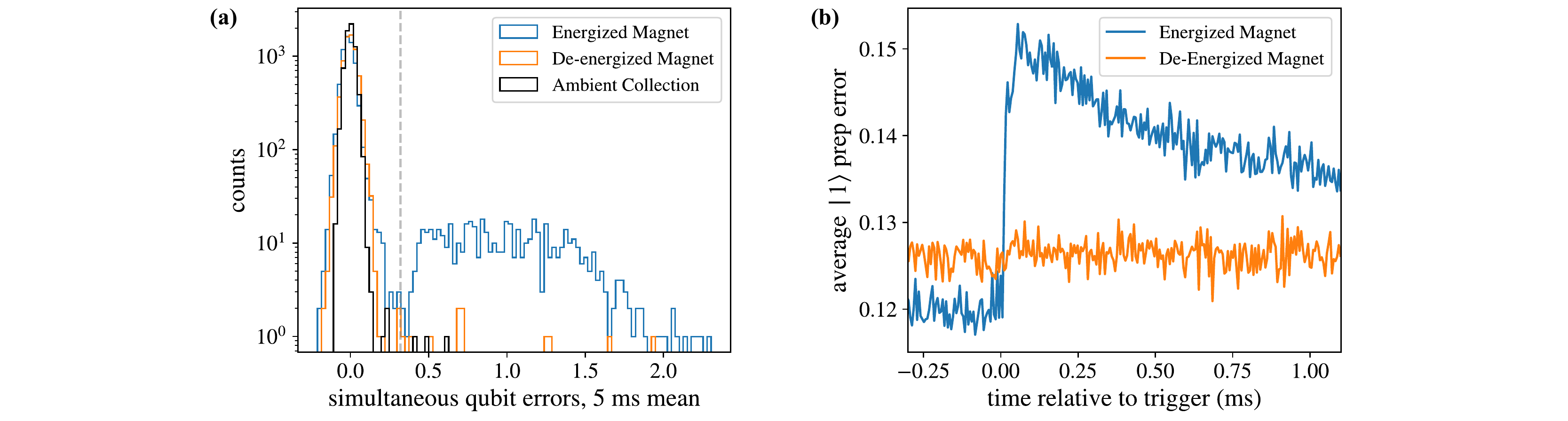}
\caption{
\textbf{Control tests.} Relaxation errors correlated with the linac firing are compared for the bending magnet energized (bent towards quantum system) and de-energized (energy deposited in nearby radiator) for a 1~\si{\micro \second} detection delay. (a)~Event classification scheme showing elimination of above-threshold events for the de-energized case with results similar to the linac completely off (data analyzed with dummy triggers to match process). (b)~Average across all linac triggers for energized and de-energized magnet cases, showing no perceptible change in errors when the electron beam is not steered towards the qubit chip.
}
\label{fig:supp_control}
\end{figure}

The linac electrons are accelerated by 2856~\si{\mega \hertz} electromagnetic fields created by a klystron amplifier with a 100~\si{\kilo \volt} electron beam. The operation of the linac could feasibly create a noisy radio-frequency and ionizing-radiation environment for these qubits and was the leading reason for locating the quantum system in the adjoining room. To confirm that the measured qubit responses were due to the accelerated electrons directly impacting the system and not other interference, control tests were performed. The electron fluence was tuned to $\sim$0.1~electrons per pulse at the qubit chip and the qubits were measured for relaxation with the bending magnet energized and de-energized ($10^8$ shots, 6840 linac triggers each). When de-energized, the electrons are not redirected towards the qubits, but instead impact a tungsten radiator directly in front of the bending magnet. The results of this comparison are shown in Fig.~\ref{fig:supp_control}(a). Using the same event classification scheme described earlier for relaxation detection, we see a nearly complete extinction of qubit errors above the detection threshold for the de-energized magnet. The few remaining events can be explained by the Bremsstrahlung radiation from the tungsten radiator. An ambient data collection, where the linac is not operated, is included to highlight the similarity to the de-energized magnet case. Furthermore, by averaging across all triggers, we observe no change in the average error rate for the de-energized case, even for a single measurement shot [Fig.~\ref{fig:supp_control}(b)]. With this, we conclude that the observed error signatures following each linac trigger is due to the impact of ionizing radiation intentionally directed at the qubit chip.

\end{document}